\documentclass[journal,onecolumn]{IEEEtran}

\usepackage{algorithm,algorithmic}
\usepackage{graphicx}
\usepackage{rotating}
\usepackage[font={small}]{caption}

\usepackage{subcaption}
\usepackage{multirow}
\ifCLASSINFOpdf
\else
\fi
\usepackage[cmex10]{amsmath}
\usepackage{authblk}

\usepackage[normalem]{ulem}

\usepackage[T1]{fontenc}
\usepackage[utf8]{inputenc}
\usepackage{authblk}

\ifCLASSINFOpdf
\else
\fi
\usepackage[T1]{fontenc}
\usepackage[utf8]{inputenc}
\usepackage{authblk}

\title{More than one Author with different Affiliations}
\author[1]{Elahe Rezaei}

\author[2]{Hafez Eslami Manoochehri}
\author[1]{Babak Hossein Khalaj}

\affil[1]{Department of Electrical Engineering, Sharif University of Technology \protect \\ 
 Email: \{e\_rezaei, khalaj\}@sharif.edu}
\affil[2]{Department of Computer Science, University of Texas at Dallas \protect \\
Email: \{hafez.eslamimanoochehri\}@udallas.edu}

\begin{document}
\title{Multi-agent Learning for Cooperative Large-scale Caching Networks}
\maketitle

\begin{abstract}
Caching networks are designed to reduce traffic load at backhaul links, by serving demands from edge-nodes. In the past decades, many studies have been done to address the caching problem. However, in practice, finding an optimal caching policy is still challenging due to dynamicity of traffic and scalability caused by complex impact of caching strategy chosen by each individual cache on other parts of network. 
In this paper, we focus on cache placement to optimize the performance metrics such as hit ratio in cooperative large-scale caching networks. Our proposed solution, cooperative multi-agent based cache placement (CoM-Cache) is based on multi-agent reinforcement learning framework and can seamlessly track the content popularity dynamics in an on-line fashion. CoM-Cache is enable to solve the problems over a spectrum from isolated to interconnected caches and is designed flexibly to fit any caching networks. To deal with dimensionality issue, CoM-Cache exploits the property of locality of interactions among caches. The experimental results report CoM-Cache outperforms base-line schemes, however at the expense of reasonable additional complexity. 

\end{abstract}

\begin{IEEEkeywords}
Multi-agent reinforcement learning, cache placement, hit ratio, large-scale caching networks.
\end{IEEEkeywords}

% \fi
\IEEEpeerreviewmaketitle

\section{Introduction}
\IEEEPARstart{W}{ith} the growth of social networks, multimedia sharing web services and specially streaming of video-on-demand contents, data traffic has increased dramatically in the past few years. It has been predicted the global internet traffic will be hundredfold by 2021 \cite{networking2016forecast}. 
%In order to handle such huge traffic rate, internet service provider companies pursue solutions to increase network capacity and drastically reduce transmission cost over the network. (mishe hazf she, yekam ezafie)
One of the promising solutions is to store contents at the network's edges close to the end users. Caching networks which reduce traffic load of the core networks by equipping storage at the edge-nodes,  improve user-perceived experience, ultimately.  
Cache placement as a classic subject addresses the question: which files from a large set of files should be cached in a limited storage, so as to reduce the main server load. One step further, server load even can be deducted more by cooperation among caches which are located nearby. However, such cooperation certainly adds more complexity to the problem, particularly, when the number of local caches increases. In other words, plausible coordination cannot be achieved without hurting tractability of the cache placement strategy.

%In this paper we consider a problem in which one central server feeds some local caches through an error-free shared link. Local caches can be representative of a number of geographically spread users at that \sout{node} region. Each local cache has a finite storage to cache a limited number of contents coming from library of central server. Each request can be responded by the following ways: a) by the current cached contents of local caches preferably of the cache originally request has come from. b) by central server through the shared link. The users' requests would be served as much as possible by help of contents have been cached earlier in storage of that local cache (a hit). The missed requests will be served such that the transmission cost (could interpret as latency or other communication costs) is minimized by exploiting cooperation with other part of networks. Then, each local caches will decide which set of files should be stored in its limited-size storage. In this modelling, all requests hit the target eventually and the propagation cost (interchangeably delay or energy consumption) response to a miss is considered, in contrast to models in [][].

%Unfortunately, cache placement has proven to be a NP-Hard problem [] and therefore many approximation algorithms have been proposed [] []. 

There have been numerous studies on cache placement problem which try to improve the performance metrics such as hit ratio.
The well-known heuristic approaches are least recently used (LRU)
and least frequently used (LFU) which perform based on recency and frequency, respectively \cite{ahmed2013analyzing}\cite{garetto2015efficient}. These traditional approaches are categorized as reactive caching approaches and neither consider the pattern of content popularity nor cooperation among caches, thereby suffer from inefficiency. In \cite{giovanidis2016spatial}, an extension of the classic LRU called spatial multi-LRU, has been introduced which investigates how to best choose the actions of update, insertion and eviction of content in multiple caches instead of single-cache. In contrast to LRU where each request can only be served by one cache (the closest one), in spatial multi-LRU, there are a set of caches for any user which can serve its request. In Multi-LRU-One, if the requested file has been found in any caches which are covering a user, only one of these caches will be updated. If the object is not found in any cache, then it is inserted only in the cache closest to the user. In \cite{giovanidis2016spatial} it has been shown that Multi-LRU has better performance than employing a regular LRU to all caches independently, but it still does not take benefit of cooperation among caches in an efficient manner.

Another category of  cache placement policies is proactive caching which estimates content request patterns first and then finds the best policy based on the estimated pattern \cite{laoutaris2005optimization}\cite{poularakis2016complexity}. Reinforcement learning was employed for cache placement in \cite{blasco2014learning} by using Multi-armed bandit (MAB) to model problem. However, due to the structure of MAB, recent requests and cooperation among caches are not exploited, incurring the cost of additional exploration steps and inefficiency.
Due to the complexity issue, there are only few studies on employment of reinforcement learning in caching. \cite{sung2016efficient} employed reinforcement in caching problem while it assumed each local cache acts individually without any explicit coordination with other agents.
In \cite{avrachenkov2017low}, a game theory based caching approach has been introduced in which only the communication between the neighboring caches is allowed. At each time step, a policy is found by forming a game among the neighboring caches. In \cite{avrachenkov2017low} each cache updates its cache content selfishly by maximizing its own hit ratio subject to the content stored by neighbouring caches. \cite{avrachenkov2017low} has shown the proposed algorithm could converge to Nash equilibrium if the content popularity has stationary profile. 
Like \cite{avrachenkov2017low}, most of the previous research have been conducted on finding a solution for cache placement assuming stationary content popularity \cite{poularakis2016caching}. However, in reality, popularity of contents could be extremely dynamic over time, specifically the popularity of videos can drop after a short period of time or it may have longer lifespan. 
%Due to the availability of inexpensive high volume storages, the time scale of cache dynamics becomes larger than the lifetime of many content which requires a dynamic popularity driven cache placement strategy.
%Traditional caching strategies are only based on instantaneous traffic demands. 
Such multi-dimensional variations in popularity patterns make cache placement problem more challenging.

In this study, we focus on decision making for multi-cache placement where caches learn environment and act cooperatively to pursue a global target. Our proposed solution, cooperative multi-agent based cache placement (CoM-Cache), is a learning approach based on multi-agent reinforcement learning (MARL) framework which takes benefit of cooperation among caches and adapts to dynamicity of popularity profile. CoM-Cache exploits the property of locality of interactions among caches \cite {oliehoek2008exploiting} to achieve a trade off between coordination and complexity without notable sacrifice in performance.

\section{Problem Statement}
\label{problem-statement}

Our model consists of a central server with $N$ files (contents) in its library. For simplicity, each content needs 1 unit of the storage. The central server is connected to local caches through error-free shared link (solid line in Fig. 1). The local caches are possibly connected together through error-free links (dashed lines in Fig. 1). 
At each time step, some files from the library are requested by the users aggregated in local caches. If one user requests a file which already has a copy in the local cache associated to that user, this copy is downloaded. Otherwise, if it is possible, other local caches will serve the request at some transmission cost such as delay. If the request still has not been served (i.e., a miss), the central server inevitably sends the file through the shared link at a higher cost. Then, a decision needs to be made as to address a critical question: which files should be cached in each local cache? This decision is made in a cooperative manner based on the previous decisions, past requests and experiences in order to optimize the performance over an infinite horizon. Note that the transmission cost of sending one file through local links is reasonably less than the shared link, since the local caches located nearby generate less delay than the distant main server. 

\begin{figure}[h]
\centering
\includegraphics[scale=0.35]{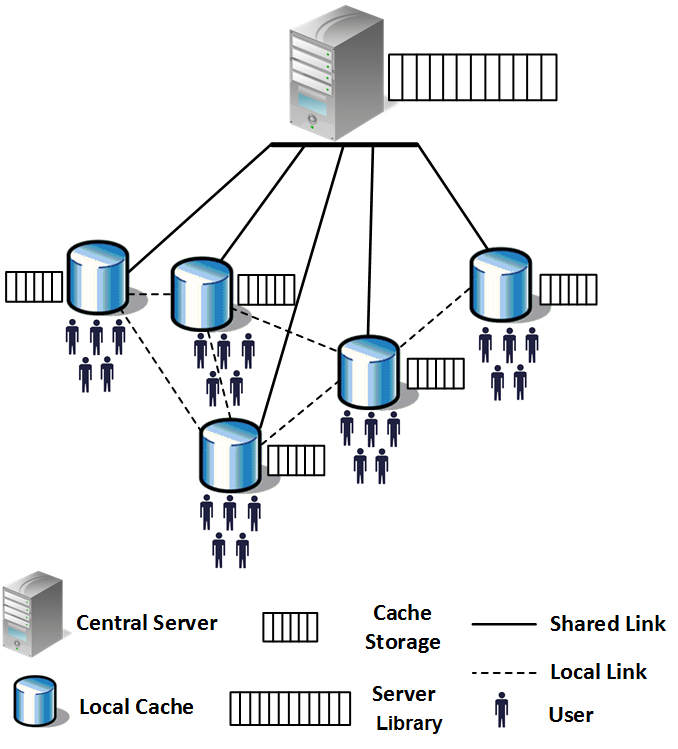}
\caption{The illustration of a caching network. Each cach is connected to a central server by a shared link and possibly connected with other caches by local links.}\label{fig:pic1}
\end{figure}

Let $M_i$ denotes the storage capacity of the $i$th local cache, $i\in {1,2,...,K}$, measured by the maximum number of files it can store.  $BW_{ij}$ denotes the capacity of link between the local caches $i$ and $j$ which indicates the maximum number of files can be transferred between $i$ and $j$ at each time step. If all $BW_{ij}s$ are set to zero, the problem is degraded into the isolated cache placement like the setup of \cite{blasco2014learning} and for $BW$ as infinity, the caching problem would be converted to the classic routing problem.
In wireless networks, the shared link and the local caches could be considered as broadcast channel and base stations of cellular networks, respectively. In this paper, the terms transmission cost and delay are used, interchangeably.

%\begin{IEEEeqnarray*}{l}
%\IEEEyesnumber \IEEEyessubnumber
%minimize \sum_t \sum_{i=1}^K \chi_i^t \IEEEyesnumber \\ 
%\IEEEeqnarraymulticol{1}{l}{\text{s.t.  } \qquad  ||\phi_i^t|| \leq M_i  \text{  }\quad \quad \; \;  \forall t, \text{   }  i \in \{1,...,K\}} \IEEEyessubnumber \\ 
%\IEEEeqnarraymulticol{1}{r}{\qquad  \quad  \; \: ||U_{ij}^t|| \leq BW_{ij}  \text{  } \quad  \forall t, \text{  } i,j \in {1,...,K}, \text{  } i \neq j}
%\IEEEyessubnumber \\
%\label{cost}
%\end{IEEEeqnarray*}
%and
%\begin{IEEEeqnarray*}{l}
%\IEEEyesnumber \IEEEyessubnumber
%maximize \sum_t \sum_{i=1}^K \theta_i^t \IEEEyesnumber \\
%\IEEEeqnarraymulticol{1}{l}{\text{s.t.  }  \qquad  ||\phi_i^t|| \leq M_i  \text{  } \quad  \quad \; \; \forall t, \text{   }  i \in \{1,...,K\}} \IEEEyessubnumber \\
%\IEEEeqnarraymulticol{1}{r}{\qquad \quad \; \: ||U_{ij}^t|| \leq BW_{ij} \text{  } \quad  \forall t, \text{  } i,j \in {1,...,K}, \text{  } i \neq j}
%\IEEEyessubnumber
%\label{hit}
%\end{IEEEeqnarray*}

%\begin{equation}
%\label{cost}
%\begin{split}
%minimize \sum_t \sum_{i=1}^K Tc_i^t \\
%\text{s.t.  } |\phi_i^t| \leq M_i  \text{  } \forall t, \text{   }  i \in \{1,...,K\} \\
%|U_{ij}^t| \leq Bw_{ij}^t  \text{  } \forall t, \text{  } i,j \in [1,...,K], \text{  } i \neq j
%\end{split}
%\end{equation}
%
%and 
%\begin{equation}
%\label{cost}
%\begin{split}
%maximize \sum_t \sum_{i=1}^K Hr_i^t \\
%\text{s.t.  } |\phi_i^t| \leq M_i  \text{  } \forall t, \text{   }  i \in \{1,...,K\}\\
%|U_{ij}^t| \leq Bw_{ij}^t \text{  } \forall t, \text{  } i,j \in [1,...,K], \text{  } i \neq j,
%\end{split}
%\end{equation}

The most popular objectives for cache placement problems are maximizing hit ratio and minimizing transmission cost (equivalently the accumulate delay).
In this paper, the optimization problem is defined as a multi-objective function. This function provides a trade-off between delay as a user-facing performance metric and hit ratio as a network level performance metric. Maximizing the hit ratio is equivalent to minimizing the rate of shared link which is defined as the ratio of total number of files sent through the shared link to the total number of requests. The constraints are: a) the storage of the local caches are limited and b) the capacity of the links between caches are limited.

\section{Cooperative Multi-agent based Cache placement}
\label{method}
In this section, we develop a mathematical framework of multi-agent decision making in CoM-Cache which is built on top of decentralized Markov decision process (Dec-MDP). 

\subsection{Dec-MDP}
A Dec-MDP is defined as a tuple $\Im=<I,S,A,T,R,h>$
where

\begin{itemize}
\item $I=\{ 1,2,...,K \}$ is a set of $K$ agents.
\item $S$ is a (finite) set of joint states.
\item  $A= \times A_i$ is the set of joint actions $a=<a_1,...,a_K>$ and $A_i$ is set of available actions to agent $i$.
\item $P$ is the transition probability function, $P: S\times A \times S \rightarrow [0,1]$ which specifies the probability of transition from state $s \in S$ to $s' \in S$ when action $a \in A$ is taken.
\item $R: S \times A \rightarrow R$ is the immediate reward function and maps states and joint actions to real numbers.
\item $h$ called horizon, is the number of steps until the problem terminates.
\end{itemize}
Each agent $i$ takes an action $a_i^t$ leading to on joint action $a^t=<a_1^t,...,a_K^t>$ at every step $t$. How the joint action influences the environment is described by the transition function $P$, called the model of the environment. When $P$ is not known, multi-agent reinforcement learning techniques are deployed to learn a solution directly without exploiting the model explicitly.
A solution to a Dec-MDP is a joint policy $\pi^*=<\pi_1,...,\pi_K>$, which is a mapping from joint states to actions and maximizes the discounted expected cumulative reward over horizon $h$. The value of a joint policy $\pi$ from state $s$ is defined as $V^{\pi}(s) = E[\sum_{t=0}^{h-1} \gamma^t R(s^t,a^t|s,\pi)]$ which represents the expected value of the reward of agents summed over time starts from state $s$ and follows policy $\pi$. In the finite-horizon case the discount factor, $\gamma$ is typically set to 1 while for infinite-horizon, $\gamma \in (0,1)$. An optimal policy beginning at state $s_0$ is defined as $\pi^*(s_0) = argmax_{\pi} V_{\pi}(s_0)$ and can be found efficiently with dynamic programming techniques.

The most common MARL algorithm is multi-agent Q-learning in which agents learn optimal mapping from the environment's states to actions when accumulative reward over time is maximized. Each state-action pair $(s, a)$ has a value called the Q-value that represents the expected long-term cumulative reward by taking action $a$ at state $s$. At each step $t$, the agents observe current states and execute actions that belong to the available set of actions $A$ and the Q-value are updated according to the immediate reward at time $t$, $r(s^t,a^t)$ as follows:
%\begin{equation}
%\label{Eq3}
%\begin{split}
%Q^k(s^k,a^k) = (1-\alpha_k) Q^{k-1}(s^k,a^k) \\ + \alpha_k [r(s^k,a^k) 
% + \gamma max_{a^{k+1} \in A} Q^{k-1} (s^{k+1}, a^{k+1})] 
%\end{split}
%\end{equation}
\begin{equation}
Q_{new}(s^t,a^t) = (1-\alpha^t) Q_{old}(s^t,a^t)+\alpha^t [r(s^t,a^t) 
 + \gamma max_{a \in A} Q(s^{t+1}, a)]
\label{qlearning1}
\end{equation}
where $\alpha^t \in (0,1)$ is referred to as the learning rate at time $t$ models the weight of learning with respect to the old information and goes to zero over time.

\subsection{CACHE PLACEMENT FORMULATION IN MARL FRAMEWORK}

At each step $t$, each cache interacts with the environment, makes a decision and executes an action. First, the users' demands can be served by the aforementioned manner, i.e. a requested content 1) would be accessed directly from a local cache (if available), 2) be obtained from one of the neighbour local caches (if possible), 3) be downloaded from central server. These ways are ordered according to their priorities. Then, each local cache updates its contents based on the learned policy. The policy determines which content should be cached where and when in order to maximize the total reward over time. 

The components of modelling cooperative cache placement problem as MARL are described as follows:

\textit{a) Set of agents:} $I$ refers to a set of local caches, $i$'s i.e. $1 \leq i \leq K$ with storage $M_i$. Thereafter, agent and cache are used interchangeably.

\textit{b) States:} $s^t$ is joint state of all local caches at time $t$, i.e. ${s}^t = (s_1^t,...,s_K^t)$. The state of agent $i$ at time $t$, $s_i^t$, is defined as a pair $[\phi_i^t,q_i^t]$ where $\phi_i^t$  and  $q_i^t$ respectively denote the set of cached files and requested files by agent $i$ during time interval $[t-1,t)$. Since the number of requested files at each time and the size of storage of one cache, never exceed the library size\footnote{If $M_i >= N$, the problem will have the trivial solution of copying all library in cache}, the space of set of states is finite. 

\textit{c) Actions:} ${a}^t= (a_1^t,...,a_K^t)$ where $a_i^t$ specifies the action of agent $i$ at time $t$. $a_i^t$  selects a set of  $M_i$ (distinct) files from $\phi_i^t\cup q_i^t$ which needs to be placed in cache $i$ at time $t$. Given the state $s_i^t$ and action $a_i^t$, $s_i^{t+1} = [\phi_i^{t+1}, q_i^{t+1}] = [a_i^t, q_i^{t+1}]$.

%\begin{equation}
%\label{Eq5}
%s_i^{t+1} = (\phi_i^{t+1}, q_i^{t+1}) = (a_i^t, q_i^{t+1})
%\end{equation}

\textit{d) Reward Function:} various forms of objective functions in cache placement problem are introduced \cite{li2016popularity}, \cite{hachem2015content}. The two popular objective functions are: a) maximizing the hit ratio and b) minimizing the transmission cost. CoM-Cache considers the linear combination of these two objectives to ensure that while the main interest of caching that is decrease traffic load at backhaul links, is achieved, the average delay experienced by the users is not overlooked either.

\subsection{Limited Interaction in Cooperative Caching}

In multi-agent systems, MARL provides a promising solution for agents which explore environment and adapt their behaviours to the dynamics of the uncertain and evolving environment. However, employing MARL for cache placement in large-scale networks comes with scalability issue. The policy space grows exponentially with the increase in population of local caches \cite{zhang2011coordinated}.
If each agent only cares about its local states and decides individually, it would result in locally optimized but not necessarily a globally optimal policy [20]. To our best knowledge, previous studies widely ignore the cooperative decision making in caching problem. CoM-Cache presents a modified collaboration model among caches where the complexity increases only polynomially with the size of network.

\textbf{Decomposition of Value Function:}
To deal with dimensionality issue, we aim to decompose the reward function into the sum of local functions over a smaller number of states and actions (smaller scope). The local reward function $R_i^t(s,a)$ is the total reward that agent $i$ can achieve by interacting with other agents. Note that local reward should not be mixed up with individual reward which refers to the reward of agent $i$ obtained selfishly without considering the global objective.

\textbf{Proposition 1} \textit{For each agent $i$ and its neighboring set $\mathcal{N}_i$ which includes all agents connected to agent $i$ thorough one-hop communication, the total reward function of agent $i$ interacting with all other caches is equal to the rewards it receives with interacting with its neighbouring set $\mathcal{N}_i$:}
\begin{IEEEeqnarray*}{l}
\IEEEeqnarraymulticol{1}{c}{R_i^t (s,a)
=R_i^t (s_i^{t},s_{\mathcal{N}_i}^{t},a_i^{t},a_{\mathcal{N}_i}^{t})}
\IEEEyesnumber
\label{local-rewards}
\end{IEEEeqnarray*}
As each agent can only receive files from one-hop neighbour via the local links, $\mathcal{N}_i$ are the only agents influence the decision making of $i$, directly. We exploit this limited dependency to shrink the scope of local reward functions.
In order to find the optimal policy, the value function which is defined over joint states and actions of all agents, needs to be computed. The value function can be decomposed into $K$ local values where each one involves only a subset of agents by factorizing the value function for decomposed immediate reward function \cite{oliehoek2008exploiting}. 

\textbf{Definition 1:} \textit{Joint probability distribution over neighbouring set $\mathcal{N}_i$ is defined as:}
\begin{equation}
P(s_i^{t},s_{\mathcal{N}_i}^{t}|s^{t-1},a^{t-1}) = \sum_{s_{I \setminus \{i,\mathcal{N}_i\}}^t}P(s_i^{t},s_{\mathcal{N}_i}^{t},s_{I \setminus \{i,\mathcal{N}_i\}}^t|s^{t-1},a^{t-1}) 
\end{equation}
where $I \setminus \{i,\mathcal{N}_i\}$ represents the set of all agents excluding $i$ and $\mathcal{N}_i$. This equations is extracted by marginalizing the joint transition probability function where $P(s^t|s^{t-1},a^{t-1})$ is written as $P(s_i^{t},s_{\mathcal{N}_i}^{t},s_{I \setminus \{i,\mathcal{N}_i\}}^t|s^{t-1},a^{t-1})$. The formulization of value function of joint policy $\pi$ started from state $s$ is then decomposed as:
\begin{equation}
V^{\pi}(s)=\sum_{i \in I}V_i^{\pi}(s)= \sum_{i \in I} \sum_{s'_i,s'_{\mathcal{N}_i}}P(s'_i,s'_{\mathcal{N}_i}|s,\pi)Q_i^{\pi}(s'_i,s'_{\mathcal{N}_i},a_i,a_{\mathcal{N}_i})
\label{1234}
\end{equation}
where $V_i^{\pi}(s) = E[\sum_t \gamma^t R_i(s^t, a^t|s,{\pi})]$ is the local value of $i$ over joint policy $\pi$ and the local Q-value over joint policy $\pi$ at time $t$ is given by: 
\begin{equation}
Q_i^{\pi}(s_i^t,s_{\mathcal{N}_i}^t, a_i^t,a_{\mathcal{N}_i}^t)= R_i(s_i^t,s_{\mathcal{N}_i}^t, a_i^t,a_{\mathcal{N}_i}^t)+\sum_{s_i^{t+1},s_{\mathcal{N}_i}^{t+1}}P(s_i^{t+1},s_{\mathcal{N}_i}^{t+1}|s^t,a^t) Q_i^{\pi}(s_i^{t+1},s_{\mathcal{N}_i}^{t+1}, a_i^{t+1},a_{\mathcal{N}_i}^{t+1})
\label{local-rewards}
\end{equation}

\textbf{Principle of Locality of Interaction:} \\
Locality of interaction has been introduced in decentralized partially observable MDP (Dec-POMDP) and has been explored in several studies \cite{melo2009learning}. However, locality of interaction relies on a strong assumption of Transition (Observation in Dec-POMDP) Independence \cite{becker2004solving}. 

\textbf{Definition 2:} \textit{A Dec-MDP is called Transition Independent (TI) if the state transition probabilities are factorized as follows:}
\begin{equation}
P(s'|s,a) = \prod_i^{K} P_i(s'_i|s_i,a_i)
\end{equation}
where $P_i(s'_i|s_i,a_i)$ represents the transition probability of agent $i$ takes action $a_i$ and transits from local state $s_i$ to $s'_i$. If we assume \textit{TI}, the value function in  Eq. \ref{1234} can be decomposed with shrunk scope. This feature which is so-called as \textit{locality of interaction} says that the local utility of agent $i$ from policy $\pi$ to $\pi'$ does not change if $\pi$ and $\pi'$ have similar mapping of states to actions for agents $i$ and the set, to which agent $i$ has interaction \cite{nair2005networked}. 
In CoM-Cache learning algorithm (see section \ref{XX}), we approximate $P(s'_i,s'_{\mathcal{N}_i}|s,a)$ by $P(s'_i,s'_{\mathcal{N}_i}|s_i,s_{\mathcal{N}_i},a_i,a_{\mathcal{N}_i})$ which shrinks the scope of value function from $(s,a)$ to $(s_i,s_{\mathcal{N}_i},a_i,a_{\mathcal{N}_i})$. The approximation comes from this observation that in caching networks, each cache has major interactions with a limited number of caches which are located in the same geographical region and have file transfer through one-hop communication.

\subsection{MARL Algorithm in CoM-Cache }
\label{XX}
The goal of each cache is computing joint policy that maximizes expected total reward of all agents. Without any coordination, agents decide based on their local observation which results into $K$ individual policies. On the other hand, the globally optimal policy that maps joint states to joint actions, inherently, performs better than individual policies. %In CoM-Cache, we assume local caches can communicate with their neighbours and they choose action, cooperatively. 
Finding the globally optimal policy requires instantaneous, loss-less and free communication therefore, it is resource demanding and infeasible in practice \cite{bernstein2002complexity}. 
CoM-Cache takes advantage of a limited interactions, which provides scalability by allowing agents to learn based on the limited but more effective observations. 
%By exploiting the principle of locality of interaction, it is possible to partition the state space to partial state spaces that consist of a few number of agents, hence the size of the partial state space is not exponentially growing with number of all agents.
By use of this fact, in CoM-Cache each local cache learns the joint policy with a set of its neighbours not the entire network. So the size of partial state space for agent $i$ is limited to $|s|^{|\mathcal{N}_i|+1}$ regardless of the size of the network $K$. 
Using the utility decomposition structure, we can define the approximate utilities such that:
\begin{equation}
\label{hat}
\hat{Q}(s,a) = \sum_i^K Q_{i,\mathcal{N}_i} ([s_i,s_{\mathcal{N}_i}],[a_i,a_{\mathcal{N}_i}])
\end{equation}
where $Q_{i,\mathcal{N}_i}$ is the utility of agent $i$ by interacting only with the set of $\mathcal{N}_i$. Note that in case of \textit{TI}, the approximate utility yields the accurate value. Clearly, the complexity of learning approximate Q-value in Eq. \ref{hat} is much less than Eq. \ref{qlearning1}.

Algorithm 1, demonstrates the learning procedure in CoM-Cache.
For agent $i$ and its neighbouring set $\mathcal{N}_i$, $\Gamma_{i,\mathcal{N}_i}^t$ is defined as the probability of taking joint action $a_{\mathcal{N}_i}$ at joint state $<s_i,s_{\mathcal{N}_i}>$ at time $t$ for any state-action pair and is calculated as follows:  
\begin{equation}
\label{pi}
\Gamma_{i,\mathcal{N}_i}^t ([s_i,s_{\mathcal{N}_i}],a_{\mathcal{N}_i}) = \frac{f([s_i,s_{\mathcal{N}_i}],a_{\mathcal{N}_i})^t}{\sum_{a'_{\mathcal{N}_i}}f([s_i,s_{\mathcal{N}_i}],a'_{\mathcal{N}_i})^t}
\end{equation}
where $f([s_i,s_{\mathcal{N}_i}],a_{\mathcal{N}_i})^t$ represents the number of observing state-action pair $([s_i,s_{\mathcal{N}_i}],a_{\mathcal{N}_i})$ during $[0,t]$. To learn the optimal joint policy, each agent $i$ needs to find the Q-values of joint policies with $\mathcal{N}_i$. The Q-value of a state-action pair $([s_i,s_{\mathcal{N}_i}],[a_i,a_{\mathcal{N}_i}])$, $Q_{i,\mathcal{N}_i}^t ([s_i,s_{\mathcal{N}_i}],[a_i,a_{\mathcal{N}_i}])$, is updated according to:
\begin{equation}
 Q_{i,\mathcal{N}_i}^{t+1} ([s_i,s_{\mathcal{N}_i}],[a_i,a_{\mathcal{N}_i}]) = (1-\alpha^t) Q_{i,\mathcal{N}_i}^{t} ([s_i,s_{\mathcal{N}_i}],[a_i,a_{\mathcal{N}_i}])+ \alpha^t[r_i^t([s_i,s_{\mathcal{N}_i}],a_i) + \gamma \Theta_i^t(s_i,s_{\mathcal{N}_i)}]
\label{update-rule}
\end{equation}

where $0 < \alpha < 1$ the learning rate and $0 < \gamma < 1$ the discount factor are preselected. $\Theta_i^t(s_i,s_{\mathcal{N}_i})$ denotes the best response for $i$ according to $\mathcal{N}_i$s' actions at time $t$ and is evaluated as follows:
\begin{equation}
\Theta_i^t(s_i,s_{\mathcal{N}_i}) = max_{a_i}[\sum_{a_{\mathcal{N}_i}}Q_{i,\mathcal{N}_i}^t ([s_i,s_{\mathcal{N}_i}], [a_i, a_{\mathcal{N}_i}]) \times \Gamma_{i,\mathcal{N}_i}^t ([s_i,s_{\mathcal{N}_i}],a_{\mathcal{N}_i})]
\label{best-response}
\end{equation}
Note that the contribution of $\Theta_i^t(s_i,s_{\mathcal{N}_i})$ to the global value might be less than $max_{a_i,a_{\mathcal{N}_i}}Q_{i,\mathcal{N}_i}^t ([s_i,s_{\mathcal{N}_i}], [a_i, a_{\mathcal{N}_i}])$, where in $\Theta_i^t(s_i,s_{\mathcal{N}_i})$ the expected value instead of maximization over $a_{\mathcal{N}_i}$ is adopted.
At each stage of the algorithm, with probability of $1-\epsilon$, the next action is found by:
\begin{equation}
a_i^{t+1} = argmax_{a_i}[\sum_{a_{\mathcal{N}_i}}Q_{i,\mathcal{N}_i}^t ([s_i,s_{\mathcal{N}_i}], [a_i, a_{\mathcal{N}_i}])\times \Gamma_{i,\mathcal{N}_i}^t ([s_i,s_{\mathcal{N}_i}],a_{\mathcal{N}_i})]
\label{decision-rule}
\end{equation}

\begin{algorithm}
\label{alg}
\caption{Learning in CoM-Cache for agent $i$}
%\begin{algorithmic}[1]
%\renewcommand{\algorithmicrequire}{\textbf{Input:}}
%\renewcommand{\algorithmicensure}{\textbf{Output:}}
%\renewcommand{\algorithmicensure}{\textbf{Initialization Phase}}
%\end{algorithmic} 
\textbf{Initialization:} \\
Let $t=0$, $\forall i \in \{ 1,..., K \}$ randomly select $s_i^t,s_{\mathcal{N}_i}^t, a_i^t, a_{\mathcal{N}_i}^t$,\\
$\Gamma_{i,\mathcal{N}_i}^t ([s_i,s_{\mathcal{N}_i}],a_{\mathcal{N}_i}) = \frac{1}{||a_{\mathcal{N}_i}||}$, $Q_{i,\mathcal{N}_i}^t ([s_i,s_{\mathcal{N}_i}],[a_i,a_{\mathcal{N}_i}]) = 0$ \\ 
\textbf{while} ($t < h$) \textbf{do}\\
\parbox[t]{2mm}{\rotatebox[origin=c]{90}{ Decision Rule }}
$
\begin{cases}
$\text{   Randomly generate } $\eta \in [0,1]$ $ \\
$\text{   \textbf{if} } $\eta \leq 1-\epsilon$ \textbf{then} $\\
$\text{   \quad Select } $a_i^{t+1}$ \text{ according to Eq. \ref{decision-rule}}$ \\
$\text{   \textbf{else} \; Randomly select } $a_i^{t+1}$ \text{ from available actions}$
\end{cases}
$

\text{ } \quad Compute the $\Gamma_{i,\mathcal{N}_i}^t ([s_i,s_{\mathcal{N}_i}],a_{\mathcal{N}_i})$ according to Eq. \ref{pi} \\
\text{ } \quad Compute the $\Theta_i^t$ according to Eq. \ref{best-response}\\
\text{ } \quad Update $Q_{i,\mathcal{N}_i}^t ([s_i,s_{\mathcal{N}_i}],[a_i,a_{\mathcal{N}_i}])$ according to Eq. \ref{update-rule}\\
\text{ } \quad $t \leftarrow t+1$ \\
\textbf{end while}
\end{algorithm}
The decision rule in algorithm 1 includes exploration phase where the agent can simply choose an action randomly. In algorithm 1, $\epsilon$ denotes an adjusting parameter to control the trade off between exploration and exploitation in learning procedure. 

% \begin{algorithm}
% \label{alg}
% \caption{CoM-Cache}
% %\begin{algorithmic}[1]
% %\renewcommand{\algorithmicrequire}{\textbf{Input:}}
% %\renewcommand{\algorithmicensure}{\textbf{Output:}}
% %\renewcommand{\algorithmicensure}{\textbf{Initialization Phase}}
% %\end{algorithmic} 
% \textit{Initialization} \\
% let $t=0$\\
%  $\forall i \in \{ 1,..., K \}$ \\
% randomly select $s_i^t,s_{O_i}^t, a_i^t, a_{O_i}^t$ \\
% $\pi_{i,O_i}^t ([s_i,s_{O_i}],a_{O_i}) = \frac{1}{||a_{O_i}||}$ \\
% $Q_{i,O_i}^t ([s_i,s_{O_i}],[a_i,a_{O_i}]) = 0$ \\ \\ 
% \textit{Learning Phase} \\
% For  $t \neq 0$ \\
%$\pi_{i,O_i}^t ([s_i,s_{O_i}],a_{O_i}) = \frac{f([s_i,s_{O_i}],a_{O_i})}{\sum_{a_{O_i}}f([s_i,s_{O_i}],a_{O_i})}$ \\
%$B_i^t = max_{a_i}[\sum_{a_{O_i}}Q_{i,O_i}^t] ([s_i^t,s_{O_i}^t], [a_i^t, a_{O_i}^t)] \times \pi_{i,O_i}^t ([s_i,s_{O_i}],a_{O_i})$ \\ \\ 
% \textit{Updating Rule of Q-value} \\
% Update $Q_{i,O_i}^t ([s_i,s_{O_i}],[a_i,a_{O_i}])$ according to eq()
% %$Q_{i,O_i}^t ([s_i,s_{O_i}],[a_i,a_{O_i}]) = (1-\alpha) Q_{i,O_i}^{t-1} ([s_i,s_{O_i}],[a_i,a_{O_i}]) + \alpha[R_i^t + \gamma B_i^t]$ \\ \\
% 
%\textit{Decision Rule} \\
%with probability $1-\epsilon$ \\
%$a_i^{t+1} = argmax_{a_i}[\sum_{a_i,a_{O_i}} Q_{i,O_i}^t ([s_i,s_{O_i}],[a_i,a_{O_i}]) \times \pi_{i,O_i}^t ([s_i,s_{O_i}],a_{O_i})]$ \\
%with probability $\epsilon$ \\
%randomly select $a_i^{t+1}$ from set of available actions.
% \end{algorithm}
 
\subsection{Complexity Discussion}
A cooperative sequential decision making in multi-agent systems can be done by Dec-MDP. However, finding the optimal solution for $h$-horizon Dec-MDP is \textit{NEXP-complete} \cite{bernstein2002complexity} and practically intractable. In the same manner, learning a globally optimal policy by MARL is not feasible when the problem scales up since the joint utility function conditions on all agents. On the other end of the spectrum, Independent Q-Learning (IQL) avoids scalability issue where each agent independently learns its own policy based on its local observation (does not condition on the state and action of other agents). IQL which was employed in \cite{sung2016efficient} cannot take advantage of cooperation among caches. For example, an oversaturated request traffic in one local cache could easily be resolved by network-wide cooperation. The number of policies to be evaluated in IQL in the general form is $O(|a|^{|s|^{h}})$ where $a$ and $s$ are the set of individual actions and states of one agent. IQL has less complexity, however suffers from poor performance in cooperative decision making \cite{claus1998dynamics}. 
In \cite{dehghan2015complexity}, the complexity of a joint problem of in-network content caching and routing, corresponds to placement and delivery phases, has been investigated where cached content can be accessed through multiple network paths. It has been proven, the jointly optimization problem is NP-complete even if there is only one local cache and each content is requested by only one user\footnote{This case is referred as congestion-sensitive delay model which quite fits with our model.}.

To avoid the computational complexity, CoM-Cache deploys one-hop communication and prioritization of serving ways of requests, which degrades the joint problem of routing and cache placement into a single optimization problem. 
On the other hand, In CoM-Cache, each agent chooses a policy considering only a certain number of agents which have the main influence through one-hop communication. So the complexity of CoM-Cache in worst case would be $O(|a|^{|\mathcal{N}|.|s|^{|\mathcal{N}|.h}})$ where $|\mathcal{N}|\ll  |I|$. 
Therefore, the computational complexity of CoM-Cache does not grow exponentially with the size of network and is comparable to IQL.

\section{Experiments}
\label{result}

In this section, we present the numerical results of the experiments
that have been conducted to evaluate the performance of CoM-Cache in compare to the benchmark caching techniques.

\subsection{Evaluation Setup} 

\subsubsection{Request Pattern}

\begin{figure*}
\centering
\begin{minipage}[b]{.45\textwidth}
  \includegraphics[width=\linewidth]{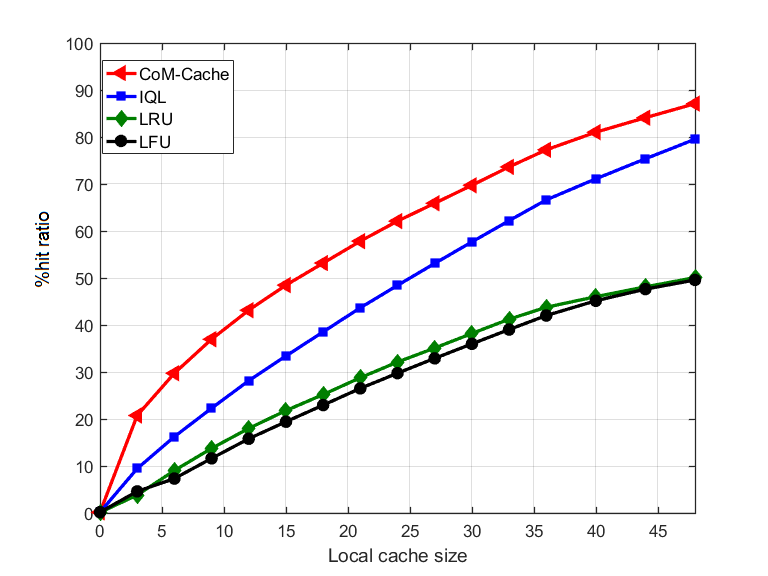}
\caption{ \%\textit{hit ratio} when request pattern is generated by IRM.}\label{cache:fig2}
\end{minipage}\qquad
\begin{minipage}[b]{.45\textwidth}
  \includegraphics[width=\linewidth]{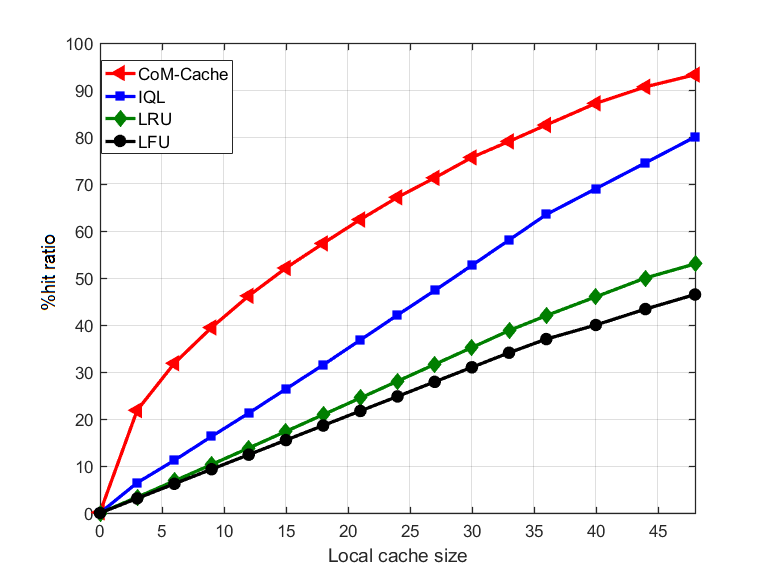}
\caption{ \%\textit{hit ratio} when request pattern is generated by SNM. }\label{cache:fig1}
\end{minipage}
%\caption{a) shows an influence graph with 4 agents. b) shows the corresponding coordinated graph.}
\end{figure*} 

Although in the literature, Independent Reference Model (IRM) model has been widely deployed for modelling content popularity  \cite{coffman1973operating}, this model ignores temporal localities and heterogeneous request distributions. To simulate the evolution of content popularity over time, we exploit Shot Noise Model (SNM) introduced by \cite{traverso2013temporal} along with Zipf distribution which generates requests as a superposition of some independent processes each one corresponding to one content. Thus, the experiment is designed for two different scenarios:
\begin{itemize}

\item  When neither temporal nor spatial correlation exists. Users request files independently from their past or their neighbours' requests (IRM). For this experiment, a Zipf distribution with the exponent parameter $\beta$, where the probability of request for the $n^{th}$ most popular file is proportional to the $n^{-\beta}$, is generated for content popularity.

\item  When temporal and spatial correlation exist. For this scenario SNM is deployed where a time inhomogeneous Poisson process describes the request pattern for content $n \in N$. In SNM, to model temporal correlation, three parameters are defined: $\tau_n$, the time instant which $n$ is requested by users for the first time, $V_n$, average number of requests generated by content $n$, and $\lambda_n(t)$, popularity profile of content $n$ over time. The idea of SNM can also be employed to  capture spatial correlation. Spatial correlation of requests comes from this fact that users from the same geographical region may have more similar taste and desire. The whole networks are partitioned in groups with few members in each one, (we select 4). A pair of $(V, \tau)$ from a joint distribution of $(V,\tau)$ is applied for all members of each group. The joint distribution of $(V, \tau)$ is selected in a way that for files with similar lifespan, the more popularity has a larger $V$.

\end{itemize}

\subsubsection{Network Setup}

We consider a caching networks consisting of a central server with local caches uniformly distributed in a square field as a grid topology. However, our work can immediately be extended to any variant of network topologies. We assume each cache is connected to at most 4 closest caches (at corners it reduces to 2 caches). The transmission ranges among caches (and interference ranges in case of wireless networks) are limited to the closest one-hop neighbours which is identical for all caches in the grid topology. For simplicity, let's assume the caches have similar storage capacities and $BW$s are identical for all links between caches.

\subsubsection{Benchmarks}

\begin{figure*}
\centering
\begin{minipage}[b]{.45\textwidth}
  \includegraphics[width=\linewidth]{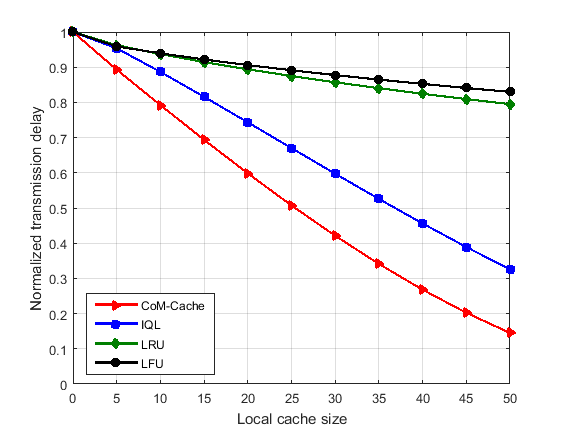}
\caption{ Normalized transmission delay when request pattern is generated by IRM.}\label{cache:fig33}
\end{minipage}\qquad
\begin{minipage}[b]{.45\textwidth}
  \includegraphics[width=\linewidth]{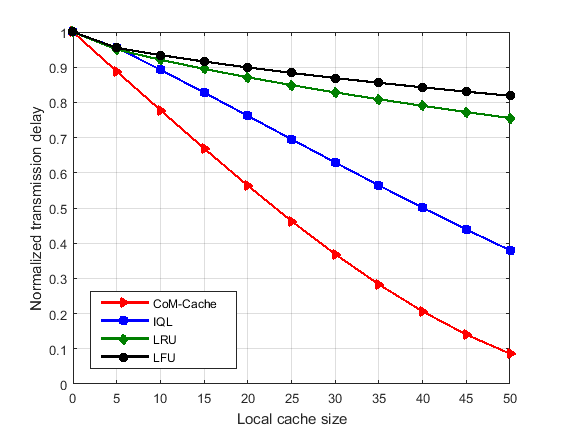}
\caption{ Normalized transmission delay when request pattern is generated by SNM. }\label{cache:fig44}
\end{minipage}
\end{figure*}

To evaluate the performance of CoM-Cache, we compare our proposed solution with the most widely adopted benchmarks, LRU and LFU caching \cite{ahmed2013analyzing}. In LRU, each local cache keeps a record of ordered list of the recent access of current cached contents. When a file which has not been cached earlier is requested, it is inserted into cache memory and the earliest file fetched from the full cache is discarded. In LFU, each local cache keeps a record of ordered list of the numbers of access of current cached contents. LFU drops a file which has the least frequency of usage over a given window in past. 
In addition to LRU and LFU, an advanced technique IQL which runs an individual Q-learning for each cache without any coordination with others, is deployed. In IQL, each cache finds the placmeent strategy only by observing its own state, and is simulated according to the setup of \cite{sung2016efficient}. 
In our experiment, the length of window for LFU and the overall horizon are set to $10^6$ and $10^{10}$ time steps, respectively.
Unless otherwise mentioned, relative cache capacity (the proportion of storage capacity in the total size of library $N) = 10\%$, $BW=1/10 \times$size of cache, exponent parameter of Zipf $\beta =0.6$ and $N=100$. In SNM, $V_n$ for the $n$th most popular file is set proportional to $1/n$, $\lambda(t)=\frac{V}\ell e^{-t/\ell}$ where $\ell$, content lifetime is uniform-randomly selected from $[10 \; 1000]$.

%We consider a network with local caches uniformly distributed in a square field and the greedy architecture in which each cache is connected to at most 4 other caches. we assume there is one central server at the center of the network as in Figure 2. 

%The optimization problem can be stated as follows: 
%At each time 
%from agent $i$ to $j$, $i$ not $j$ and $I$ form 1 to $K$. due to the symmetry of down and upstream channels, $wij$ could be considered equal  to $wji$ for all $i$ and $j$.  $L_{ij}$ is a capacity limit corresponds to the local link between $i$ and $j$ where by similar reasoning, $L_{ij} = L_{ji}$.

%For Popularity distribution of files, which is so-called popularity of reference in web caching problems, the Independent Reference Model (IRM) had been introduced by Coffman \cite{coffman1973operating}. This model assumes the request sequence as a sequence of independent random variables with the fixed stationary distribution over $N$ files. Large amount of previous research are based on IRM where it is unrealistic for dynamic environment.

\subsection{Numerical Evaluation}

%\begin{figure}[h]
%\centering
%\includegraphics[scale=0.35]{fig2.png}
%\caption{The illustration of cooperative caching. Each cach is connected to a central server by a shared link and possibly connected with other caches by local links.}\label{fig:fig2}
%\end{figure}
%
%\begin{figure}[h]
%\centering
%\includegraphics[scale=0.35]{fig1.png}
%\caption{The illustration of cooperative caching. Each cach is connected to a central server by a shared link and possibly connected with other caches by local links.}\label{fig:fig1}
%\end{figure}

We first compare the performance when the storage of caches increases for constant library size. Fig. \ref{cache:fig2} and \ref{cache:fig1} illustrate the \textit{hit ratio} achieved by CoM-Cache, IQL, LRU and LFU for different cache sizes, where \textit{hit ratio} is calculated as the percentage of requests served by local caches.
As expected, having the larger cache storage provides more possibility to serve requests by local caches.  
To investigate the impact of different traffic patterns, we run the experiment for two different request patterns. In Fig. \ref{cache:fig2}, request pattern is generated as IRM, following Zipf distribution while Fig. \ref{cache:fig1} generates requests based on SNM described before. By comparing Fig. \ref{cache:fig2} and \ref{cache:fig1}, it is understood that capturing locality in requests has noticeable impact on caching performance. For example, for size of cache $30$, CoM-Cache outperforms IQL by $12\%$ improvement in \textit{hit ratio}, while if locality between requests exists, such improvement even increases to $25\%$. This can be explained by the fact that spatial correlation of requests forces neighbouring nodes to cache similar contents in IQL, causing worse performance than cooperative learning. On the other hand, in CoM-Cache, consequent requests at the same region will largely be served by local caches rather than central server that leads to lower server load.
According to Fig. \ref{cache:fig2} and \ref{cache:fig1}, LFU yields worse long-term performance than LRU. This may happen due to the problem so called \textit{cache pollution}. This phenomenon occurs when LFU keeps a previously popular file for a long time which recently becomes unpopular and causes poor performance specially in highly dynamic environment.

In Fig. 4 and 5, the transmission delay for the same experiment are measured. The normalized transmission delay is obtained by dividing the delay of serving one request by the maximum possible delay (downloading file from the central server). Similarly, CoM-Cache provides less delay in compare to other techniques, particularly when the request pattern has temporal and spatial correlation.

%\begin{figure}[h]
%\centering
%\includegraphics[scale=0.55]{f33.png}
%\caption{\textit{Normalized transmission delay} when request pattern is generated by IRM.}\label{cache:fig33}
%\end{figure}
%
%
%
%\begin{figure}[h]
%\centering
%\includegraphics[scale=0.55]{f44.png}
%\caption{\textit{Normalized transmission delay} when request pattern is generated by SNM.}\label{cache:fig44}
%\end{figure}

\begin{figure}[h]
\centering
\includegraphics[scale=0.42]{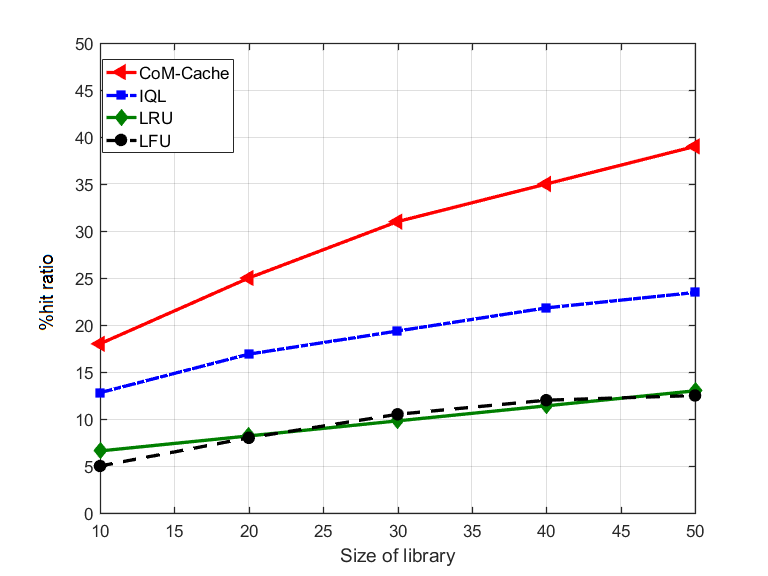}
\caption{\%\textit{hit ratio} for different size of library with relative cache capacity $= 10\%$.}\label{cache:fig5}
\end{figure}

%Further, it demonstrates that for a given cache size and pattern request, the performance of CoM-Cache is noticeably better than independent learning case. This advantage increases when the requests have correlated pattern. 

In addition to the popularity profile of contents, another important factor is the number of contents in the library. Fig. \ref{cache:fig5} complements previous experiment reporting the results obtained, where the size of library varies for fixed relative size of cache. Fig. \ref{cache:fig5} compares \textit{hit ratio} for different schemes. In all schemes, we achieve performance improvement for higher size of library. The storage of caches increasing proportionally with the number of contents, provides the larger size of total cache in network. This gain (which comes with higher cost of computational complexity) is obtained as the number of request of contents increases at a sub-linear rate compared to the number of contents, and allows caches to manage their capacity more effectively. Another interesting observation is that as the number of contents increases, the performance difference between CoM-Cache and IQL becomes larger. We can infer when the storage is limited, the role of cooperation among caches becomes more important for larger library.

%In Fig \ref{cache:fig4}, we explore the impact of changing the bandwidth of local links among caches on two metrics, hit ratio and rate of shared link for two different reward functions. The cost of local links, $w$ is set to zero and 1 in CoM-Cache I and CoM-Cache II, respectively. The average rate of shared link is defined as the ratio of number of files sent through the shared link to the total number of requests. Note that the hit ratio of LRU algorithm is unaffected by variation of $BW$, since it ignores the transmission capability among caches. Interestingly, in Come-Cache I, hit ratio is not always increasing for higher $BW$. In other word, in this case, the obtained reward is not responsive to cost of transmission between local caches, so it tries to serve requests as much as possible either by its own cache (hit) or by assist of other caches. In contrast, CoM-Cache II cares about the transmission costs so although it is most efficient in terms of hit ratio, it would not necessarily result in the minimum rate of shared link. This experiment illustrates the importance of definition of reward function and selection of parameters to control the performance.

\begin{figure}[h]
\centering
\includegraphics[scale=0.57]{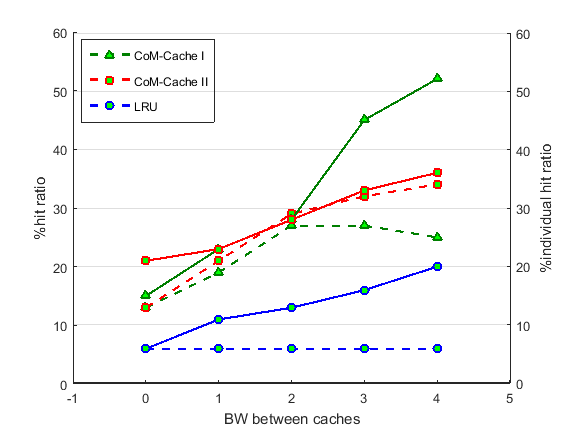}
\caption{ \%\textit{hit ratio} (shown by solid lines) and \% individual hit ratio (shown by dashed lines) over different bandwidths of local links.}\label{cache:fig4}
\end{figure}

In Fig \ref{cache:fig4}, we explore the impact of changing the bandwidth among caches on two metrics: \textit {Individual hit ratio} and \textit{hit ratio} for two different cases. \textit{Individual  hit ratio} is defined as the ratio of number of requests of one cache served by itself (not with help of the other caches) to the total number of its requests. In CoM-Cache I, we assume the communication between caches is free while in CoM-Cache II, the cost of local links is non-zero. While \textit{hit ratio} is an appropriate indicator to evaluate the performance of network of caches in cooperative sense, \textit{individual hit ratio} measures the effectiveness of individual cache placement strategy. Interestingly, in CoM-Cache I, \textit{individual hit ratio} is not always increasing for higher $BW$ although \textit{hit ratio} has fairly well increasing trend. In this case, the obtained reward is not responsive to cost of transmission between local caches, so it serves requests as much as possible by assist of other caches. In contrast, CoM-Cache II which cares about the transmission costs, has worse \textit{hit ratio} even though its \textit{individual hit ratio} surpasses CoM-Cache I. Note that the \textit{individual cache hit ratio} of LRU algorithm is unaffected by variation of $BW$, since it ignores the transmission capability among caches.

\begin{figure}[h]
\centering
\includegraphics[scale=0.38]{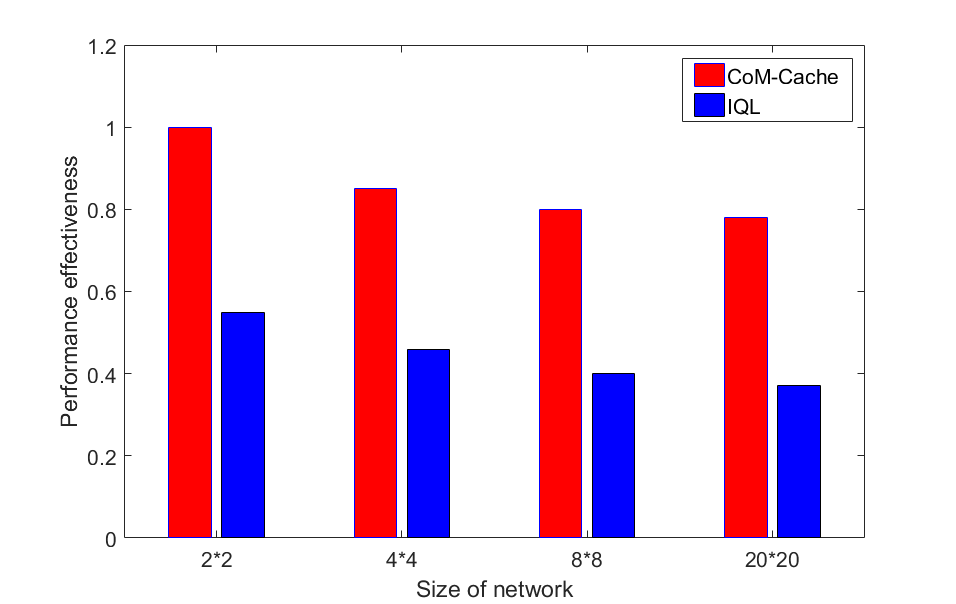}
\caption{Performance effectiveness for different size of network based on IO-UB.}\label{effec}
\end{figure}

To better evaluate the effectiveness of learning algorithm of CoM-Cache which employs the approximate decomposed utility functions, we utilize an idea introduced in \cite{oliehoek2015influence} to estimate an influence optimistic upper bound (IO-UB) in multi-agent planning problem. The idea is to compute an upper bound by relaxing the original problem with respect to the positive
impact that the rest of the network would have.
To find IO-UB in cooperative caching problem, we divide the network into sub-networks with 4 caches by eliminating some links. For any removed link, we increase the storage capacity of the two disjointed caches to the sum of capacities. This modification is optimistic since it assumes that each local cache entirely uses the maximum capacity of its neighbor. 
Thus, the original large-scale problem is converted
to non-overlapping sub-problems. In Fig. \ref{effec}, the performance effectiveness of CoM-Cache and IQL for different size of networks is demonstrated. The performance effectiveness of one scheme is defined as a ratio of \textit{hit ratio} of that scheme to \textit{hit ratio} of IO-UB. In this problem, we consider $2\times 2$, $4\times 4$ , $10\times 10$ and $20\times 20$ grid topologies where each cache is connected to at most 4 nearest neighbours. As the size of network and thus the number of removed links increases, the performance effectiveness drops. However, this reduction is not growing with the number of caches so the effectiveness of large size network e.g. $100\times 100$ would be roughly similar to the smaller network like $4\times 4$ which is about $80\%$.

\section{Conclusion}
\label{conc}

In this paper, the cooperative cache placement for large-scale caching networks has been addressed. By considering interactions among caches, a new learning algorithm, CoM-Cache, is presented which limits the scope of value functions to the neighboring set. The learning algorithm which is executed in an on-line fashion, can track the variations of traffic demands seamlessly. It is shown CoM-Cache improves both user and network level metrics, delay and hit ratio, over base-line schemes such as LRU and independent Q-learning at the reasonable cost of computational complexity. As numerical results report, CoM-Cache by taking advantage of cooperation among caches, achieves over $40\%$ reduction in server load even for small relative size of cache $10\%$.

%\input{additional}

%\section*{Acknowledgment}

\bibliographystyle{IEEEtran}
\bibliography{IEEEexample,bibliography2}

\end{document}